\documentclass[journal,letters]{IEEEtran}

\usepackage[final]{graphicx}
\usepackage[reqno]{amsmath}
\usepackage{amssymb}
\usepackage{amsthm}
\usepackage{subfig}
\usepackage{epstopdf}
\usepackage{algorithm}
\usepackage{algorithmicx}
\usepackage{algpseudocode}
\usepackage{setspace}
\usepackage[T1]{fontenc}
\usepackage{color,soul}
\usepackage{tabularx}
\makeatletter
\newcommand{\multiline}[1]{%
  \begin{tabularx}{\dimexpr\linewidth-\ALG@thistlm}[t]{@{}X@{}}
    #1
  \end{tabularx}
}
\makeatother
\algnewcommand{\LeftComment}[1]{\Statex \(\triangleright\) #1}
\usepackage{microtype}
\newtheorem{thm}{Theorem}
\newtheorem{cor}{Corollary}
\newtheorem{lem}{Lemma}

\begin{document}

\title{A Number Theoretic Approach for Fast Discovery of Single-Hop Wireless Networks}

\author{\large{Tolunay Seyfi, {\em Student Member, IEEE}}, \large{Ahmed P. Mohamed, {\em Student Member, IEEE}}, \large{Aly El Gamal, {\em Senior Member, IEEE} \vspace{-20pt}}
\thanks{The authors are with the School of Electrical and Computer Engineering of Purdue University, West Lafayette, IN (e-mail: \{tseyfi, mohame23, elgamala\}@purdue.edu).}}

\maketitle

\begin{abstract}
Interference management has become a key factor in regulating transmissions in wireless communication networks. To support effective interference management schemes, it can be essential to have prior knowledge about the network topology. 
In this paper, we build on existing results in the literature on the simulation of the message passing model, and present an efficient strategy for fast discovery of the network topology during a pilot communication phase. More precisely, we investigate the minimum number of communication rounds that is needed to discover an arbitrary network topology with a maximum number of links per receiver, while assuming a single-hop network that is restricted to interference-avoidance based schemes in its pilot phase. We first ignore any interference cancellation strategy such that no receiver can recognize, and cancel transmissions of, previously discovered transmitters, and then capture the gains obtained through interference cancellation during the pilot phase. Our results evince how the required number of rounds scale in an approximately logarithmic fashion with practical values of the total number of users in the network, having a slope proportional to the number of interfering transmitters per receiver. 


\end{abstract}

\section{Introduction}

With the increasing demand on wireless networks having strict Quality of Service (QoS) requirements, as well as the rapidly increasing network sizes, specially with the emergence of Internet of Things (IoT) applications, managing interference to deliver satisfactory performance has become more challenging. However, most effective interference management schemes that deliver significant and scalable performance gains require at least knowledge of the network topology, if not also the channel state information. At the same time, the feasibility of cloud-based centralized control of large network nodes offer unprecedented interference management opportunities \cite{ic-book}. In this work, we capitalize on this new paradigm to introduce a novel algorithm for fast discovery of the network topology, as well as making the channel state information available at the receivers. The proposed approach relies on a well-founded number-thereotic approach that enables interference-avoidance-based network discovery to complete in a number of communication rounds that scales logarithmically with the network size. 

It is important to note that even though network discovery is primarily useful for interference management, it can also be essential for core network tasks such as fault management \cite{cce}. Further, it can be particularly useful in Multiple Radio Access Technology (Multi RAT) infrastructural networks, as is the case with the cellular/WIFI network considered in \cite{lcn}.

The network discovery problem has been typically studied in a Device-to-Device (D2D) setting that allows devices to communicate directly without using infrastructural base stations. The infrastructure has been mainly considered to assist D2D network discovery, as in \cite{icufn}. In \cite{ieee-transsignal}, the network discovery problem was considered from a graph-theoretic perspective. A protocol was then introduced, capitalizing on channel sensing, random backoff, and tackling the hidden terminal interference issue, and probabilistic guarantees were provided for complete discovery of the interference graph.

In this work, we restrict our attention to a single-hop network with $K$ transmitter/receiver pairs, where a coordinated interference-avoidance scheduling strategy is orchestrated by a cloud-based controller to enable each receiver to discover up to $L$ transmitters that could be connected to it. This considered model draws its roots from the information-theoretic $K$-user binary interference channel model with path loss constraints, as well as recent advances rendering the feasibility of cloud-based wireless communications. We rely on a well-investigated property of prime number residuals to design the proposed scheduling strategy, such that after a number of communication rounds with asymptotic complexity of $O\left(\frac{L^2 \log^2 K}{\log L \log K}\right)$, network discovery is guaranteed to successfully complete, regardless of the particular connectivity pattern. Further, provided empirical evidence suggests that for practical values of $K$ and $L$, the needed number of communication rounds scales linearly with $\log K$ with a slope proportional to $L$. We believe that this is a promising result that opens the door for supporting new applications via next generation cloud-based wireless networks.   

\section{Problem Setup}
We now elaborate the setup for the considered network discovery problem. We consider a bipartite network with $K$ transmitting (Tx) and $K$ receiving 
(Rx) nodes, where each Tx has only knowledge about its own index, and each Rx is connected to at most $L< K$ arbitrarily chosen transmitters. 
Every transmitted message carries the index of the originating Tx. During each communication round, a message can be successfully delivered from Tx $i$ to Rx $j$ if and only if no other transmitter connected to Rx $j$ is active, or it is the case that interference cancellation is employed and Rx $j$ have already discovered all other transmitters that are active in the considered round. 
Whenever an Rx decodes a message, it knows the index of the originating Tx. 
We are interested in the the minimum number of communication rounds that is needed to fully discover the network topology with and without interference cancellation, and the associated transmission scheduling strategy. 

\section{Supporting Theory and Proposed Algorithm}{
\label{sec3}
In this section, we provide the theoretical foundation for our proposed network discovery scheduling algorithm. We start with the following two lemmas, which aid in the analysis of our proposed algorithm. It is important to note that Lemma \ref{lem1} below was previously used in \cite{complexity-radio} for a single round simulation of the message passing model.
\begin{lem}
\label{lem1}
Given $s$ distinct integers $1\leq x_1,...,x_s\leq n$, there exists a prime $p_i \leq s \log_2{n}$ for each $1\leq i \leq s$ such that $x_i\not\equiv x_j\mod{p_i}$,~ $\forall j\neq i$.
\end{lem}
\begin{lem}
\label{lem2}
For any integer $x$, let $\pi(x)$ denote the number of prime numbers upper bounded by $x$, then the following holds.
\begin{equation}\label{eq:pix}
\pi(x) \leq \frac{x}{\ln x} \left(1+\frac{1.2762}{\ln x}\right).
\end{equation}
\end{lem}
\begin{IEEEproof}
The proof follows in a straightforward manner from the results stated in \cite{rosser62} and \cite{dusart99}.
\end{IEEEproof}
Inspired by the above lemmas, we can now describe the proposed algorithm. The algorithm simply consists of $n$ phases, where $n$ is the index of the largest prime number that does not exceed $L \log_2 K$. Each phase consists of $p_i$ communication rounds, where $p_i$ is the $i^{th}$ prime number with respect to ascending order. In each phase, transmitter $j$ is active only in round number $j\mod p_i$.

A network discovery simulation of the proposed algorithm is outlined in Algorithm \ref{alg:main}, where a network topology realization with connectivity parameter $L$ and $K$ users is input as a binary matrix $H$. 
It starts with generating a set of all potential communication phases, corresponding to the primes $\mathcal{P}=\{p_1,...,p_n\}$ in the range $\{2,3,...,L\log_2{(K)}\}$. A transmission schedule is then set and stored in matrix $S$ according to the above described algorithm. The reduced network topology, that consists only of the links corresponding to the active transmitters, is considered in each round, and denoted by the binary matrix $H_{active}$. Then, an active link is marked as discovered in a round, whenever it is the only active link connected to the designated receiver in that round. If receiver-side interference cancellation is enabled, then the topology that is considered at the beginning of each following round will be reduced by all the links that have already been discovered. Once the whole network topology for the given channel realization is discovered, the number of communication rounds is determined as 
\begin{align}
r=\sum_{x=1}^{m-1}(p_x)+(i\mod{p_m})+1,
\end{align}
that is the sum of rounds for all phases prior to the termination phase, plus the index of the round $(i \mod{p_m})$, when the last unknown link attached to transmitter $i$ is discovered in the $m^{\text{th}}$ phase, and a $1$ is added since the lowest mapped round has index $0$.

We then provide the following theorem to characterize the guaranteed convergence of the proposed algorithm.


\begin{thm}
\label{thm1}
For any single-hop bipartite network that is formed by $K$ transmitting and $K$ receiving nodes, where each receiving node is only connected to L transmitting nodes, the network discovery algorithm outlined in Algorithm 1 guarantees that the number of communication rounds needed to determine the network topology is bounded by $\frac{L^2 \log_2^2 K}{\ln \left(L \log_2 K\right)}\left(1+\frac{1.2762}{\ln\left(L \log_2 K\right)}\right)$.
\end{thm}


	


\begin{algorithm}[!]
\caption{Network Discovery Simulation of Proposed Algorithm}
\setstretch{1.13}
    \hspace*{\algorithmicindent} \textbf{Input} Connectivity parameter $L$, Network size $K$, Network \phantom . \phantom a \phantom .topology realization with parameters $L$ and $K$.\\
    \hspace*{\algorithmicindent} \textbf{Output} Number of communication rounds $r$.\\
     \hspace*{\algorithmicindent} \textbf{Variables} $H_{check}$: Binary matrix outlining discovered \phantom a \phantom . \phantom . \phantom . \phantom a \phantom . \phantom . \phantom . \phantom a \phantom . \phantom . \phantom . transmitting links so far.\\
     \phantom a \phantom . \phantom . \phantom . \phantom a \phantom . \phantom . \phantom a \phantom a $H_{active}$: Binary matrix outlining active \phantom a \phantom . \phantom . \phantom . \phantom a \phantom . \phantom . \phantom . \phantom a \phantom . \phantom . \phantom .\phantom a \phantom . \phantom .  transmitting links in current round.\\
     \phantom a \phantom . \phantom . \phantom . \phantom a \phantom . \phantom . \phantom a \phantom a  
     $H$: Binary matrix outlining all \phantom a \phantom . \phantom . \phantom . \phantom a \phantom . \phantom . \phantom . \phantom a \phantom . \phantom . \phantom . \phantom . \phantom . \phantom . \phantom . \phantom . \phantom .\phantom .\phantom .\phantom .\phantom .\phantom .\phantom .\phantom .\phantom .existing links in considered network topology.\\
     \phantom a \phantom . \phantom . \phantom . \phantom a \phantom . \phantom . \phantom . \phantom a \phantom . $S$: Binary matrix outlining transmission schedule 

 \begin{algorithmic}[1]
    \label{alg:the_alg}
\State{Generate the totally ordered set of primes $\mathcal{P}\gets \{2,...,p_n\}$,~ $ \max{(\mathcal{P})}~ \leq L\cdot \log_2{(K)}$;}
\\\textbf {INIT} $r \gets 0$, $H_{check} \gets 0_{K,K}$, $S \gets 0_{n,K}$;
\For{$i \gets  1$ to $n$}
\For{$j \gets  1$ to $K$}
\State Define entries $S_{i,j} \gets  j \mod {p_i}$;
\EndFor
\EndFor

\For{$m \gets 1$ to $n$} 
\For{$i \gets 0$ to $max$ ($S_{m,1},S_{m,2},...,S_{m,K}$)}
\If{Interference cancellation is enabled}
\State \multiline{%
 $H_{active}  \gets$ $H$ - $H_{check}$}
\Else{}
\State \multiline{%
 $H_{active} \gets$ $H$}
\EndIf
\For{$l \gets 1$ to $K$}, 
\If{$S_{m,l} \neq i$}
\State{$H_{active_{t,l}} \gets 0$ \textbf{ for all } $t \in \{1,2,...,K\}$};
\EndIf
\EndFor
\For{$j \gets  1$ to $K$}
\State{ \Comment search for receivers that have only one \phantom a \phantom . \phantom . \phantom . \phantom a \phantom a \phantom . \phantom . \phantom .    connection after ignoring the inactive transmitters}
\If {$(\sum_{x=1}^{K} H_{active_{j,x}})==1$}  
\State {Find $c_2$ s.t. $H_{active_{j,c_2}}==1$;}
\State { $H_{{check}_{j,c_2}} \gets  1$;}
\EndIf
\EndFor

\If{ $H==H_{check}$} \algorithmiccomment{All active links are \phantom a \phantom . \phantom . \phantom . \phantom a \phantom a \phantom . \phantom . \phantom . \phantom a \phantom a \phantom . \phantom . \phantom . \phantom a \phantom a \phantom . \phantom a \phantom a \phantom a \phantom a  \phantom . \phantom .  found}
\State{Calculate number of communication rounds \phantom a \phantom . \phantom . \phantom . \phantom a \phantom a \phantom . \phantom . \phantom .   $r \gets  \sum_{x=1}^{m-1}(p_x)+i+1$;} 
\State{break};
\EndIf
\EndFor
\EndFor
	
\end{algorithmic}
\label{alg:main}

\end{algorithm}

\begin{IEEEproof}
Applying Lemma \ref{lem1} with values $s=L$ and $n=K$ implies that for every receiver connected to $L$ transmitters, there exists a prime number (phase) $p_j \leq L \log_2 K$ for each transmitter $j$ connected to it, where transmitter $j$ can successfully deliver its message to the targeted receiver in round number $j \mod p_j$. Hence, using the proposed algorithm, network discovery is guaranteed to complete in $\pi(L \log_2 K)$ phases, where $\pi(x)$ is the number of prime numbers upper bounded by $x$. The theorem statement hence follows by applying Lemma \ref{lem2} with $x=L \log_2 K$ and noting that the $i^{th}$ phase consists of at most $p_i \leq L \log_2 K$ communication rounds. 
\end{IEEEproof}
\begin{cor}
The bound in Theorem \ref{thm1} implies an asymptotic communication complexity for single-hop network discovery of $O\left(\frac{L^2 \log^2 K}{\log L \log K}\right)$.
\end{cor}

We next discuss the simulation results for the proposed algorithm in Section \ref{sec:4} and compare the cases with and without interference cancellation.
In addition, we will examine the effect of deep fading, where each link in the network is subject independently to erasure with probability $p=\frac{1}{2}$. 
}
\section{Empirical Evaluation}{
\label{sec:4}
We perform Monte-Carlo simulations for a single-hop networks with different connectivity parameters $L\in\{3,5,7\}$ and numbers of transmitter-receiver pairs $K=2^n, n\in\{3,4,..,13\}$. The source code for the considered simulations is publicly available\footnote{https://github.com/toluhatake/A-Number-Theoretic-Approach-for-Fast-Discovery-of-Single-Hop-Wireless-Networks}.  Because of computational constraints, our simulations for each parameter setting is based on only 100 network realizations. 
We now discuss the results of the simulation for the proposed network discovery algorithm with and without interference cancellation. Figure \ref{fig:rounds} depicts the number of communication rounds - averaged over 100 realizations - to discover the network topology for different network sizes $K=2^n, n\in\{3,4,..,13\}$. Note that the axis for the network sizes represents a logarithmic scale.
We observe how the number of rounds grows in an approximately linear fashion with $\log_2(K)$ while the connectivity parameter $L$ determines the slope of that linear growth. 
Further, we observe how exploiting knowledge of already discovered links via interference cancellation at the receivers leads to significant reduction in the slope of the linear growth of the number of rounds with the logarithm of the number of users, and these two observations hold for all tested values of $L\in\{3,5,7\}$, which are of practical significance in the context of cellular networks. 
As an example for the savings introduced by using the proposed algorithm versus a time sharing scheme that requires $K$ rounds, we note that for $K=8192$ and $L=7$, an average of $180.27$ rounds are needed without interference cancellation. Note that this number of rounds is far less than the upper bound provided by Theorem \ref{thm1} for this case, which suggests the potential for further tightening of this bound, and that the proposed algorithm delivers superior performance to the provided guarantee for practical values of $L$ and significantly large values of $K$. Further, using interference cancellation reduces the average number of rounds to $76.23$ for this case. Finally, we present results with probability $\frac{1}{2}$ independent link erasures in Figures \ref{fig:fading} and \ref{fig:fading_ic} without and with interference cancellation, respectively. It is worth noting how the sparser interference graph due to link erasures results in a change in slope of the linear growth of the number of rounds with $\log K$ without interference cancellation, and in an offset with interference cancellation; both effects being more pronounced with larger values of the connectivity parameter $L$. Following the above mentioned example with $K=8192$ and $L=7$, without interference cancellation, the average number of needed rounds reduces from $180.27$ to $121.64$ due to erasures, and with interference cancellation, reduces from $76.23$ to $57.99$ due to erasures.

\begin{figure}[H]
\centering
	\includegraphics[scale=0.18]{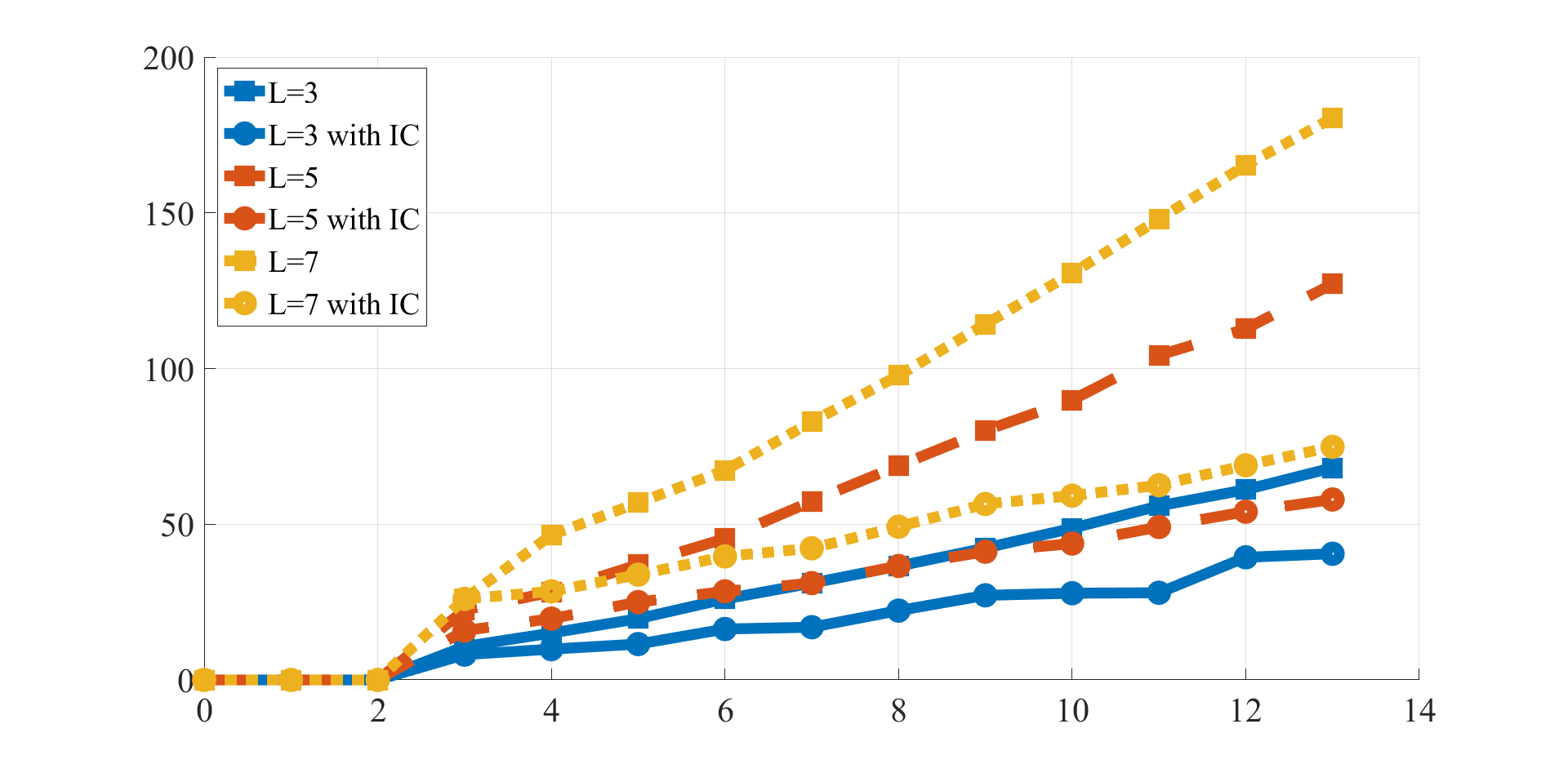}
	\caption{Number of communication rounds per $\log_2{(K)}$ with and without interference cancellation (IC).}
	\label{fig:rounds}
\end{figure}

\begin{figure}[htb]
\centering
	\includegraphics[scale=0.18]{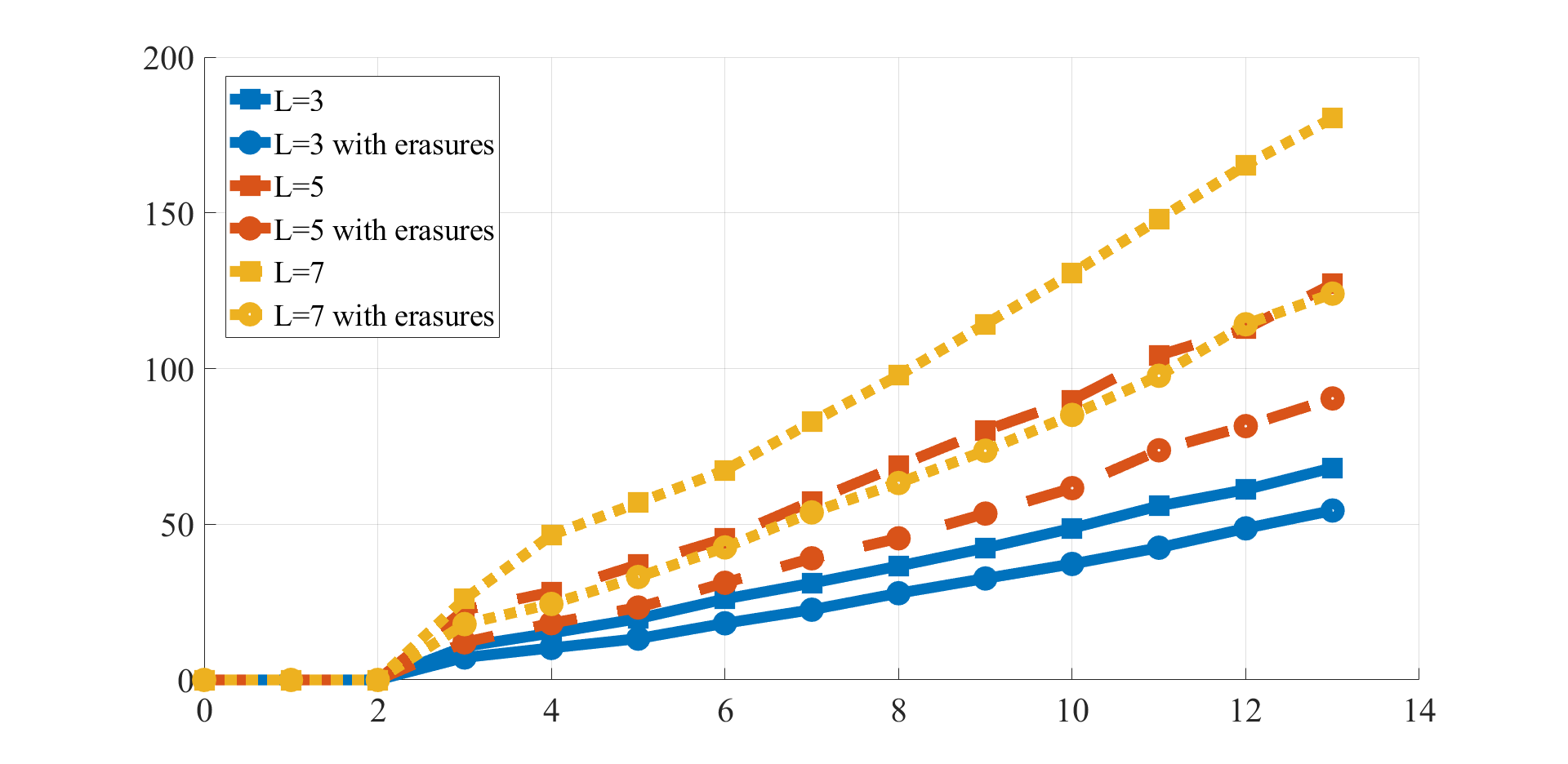}
	\caption{Number of communication rounds per $\log_2{(K)}$ with and without random link erasures.}
	\label{fig:fading}
\end{figure}

\begin{figure}[htb]
\centering
	\includegraphics[scale=0.18]{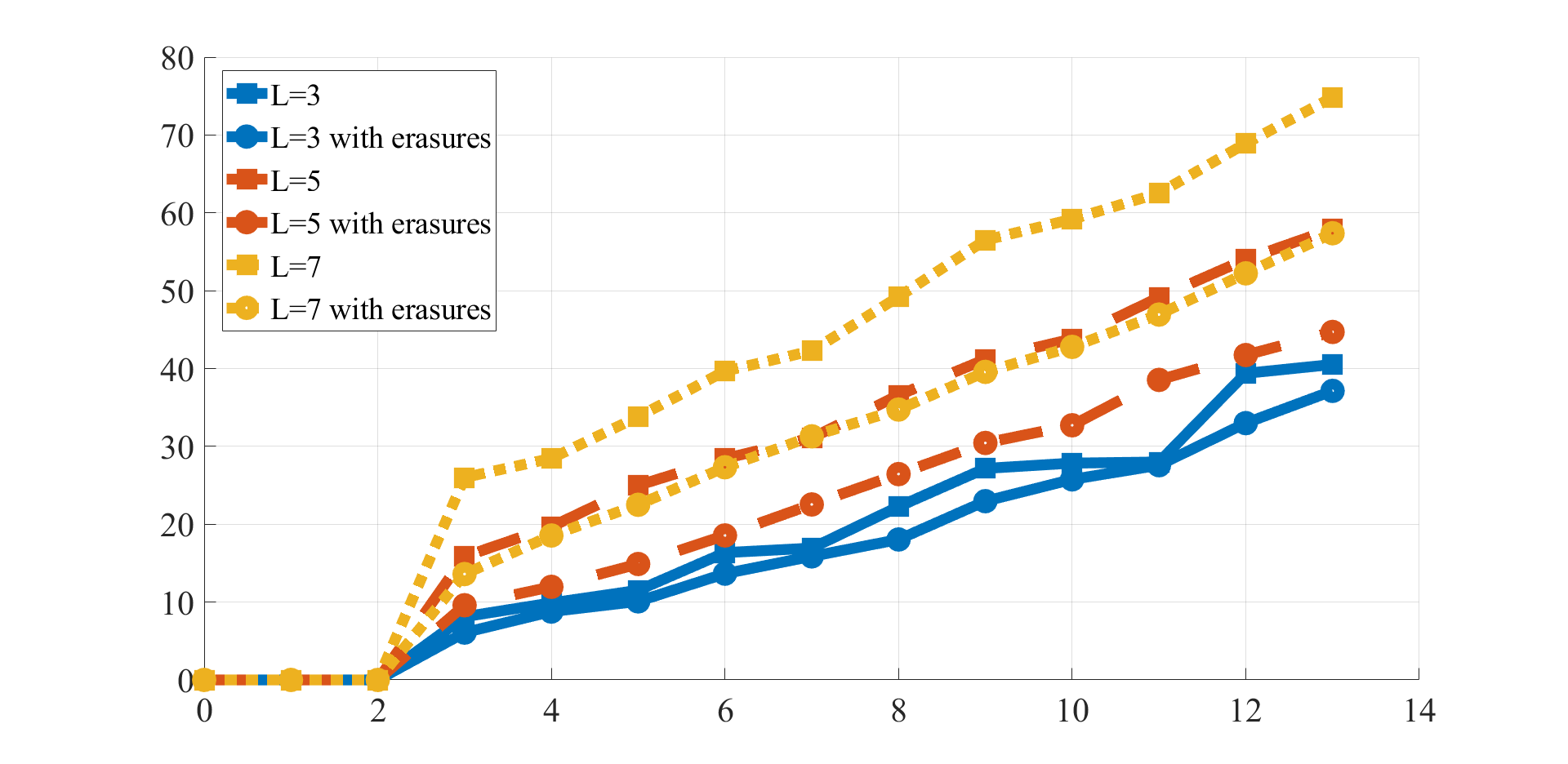}
	\caption{Number of communication rounds per $\log_2{(K)}$ with interference cancellation and with and without random link erasures.}
	\label{fig:fading_ic}
\end{figure}

\begin{figure}[htb]
  \centering 
\subfloat[]{\label{fig:l3}\includegraphics[height=0.3\textwidth, width=.2\textwidth]{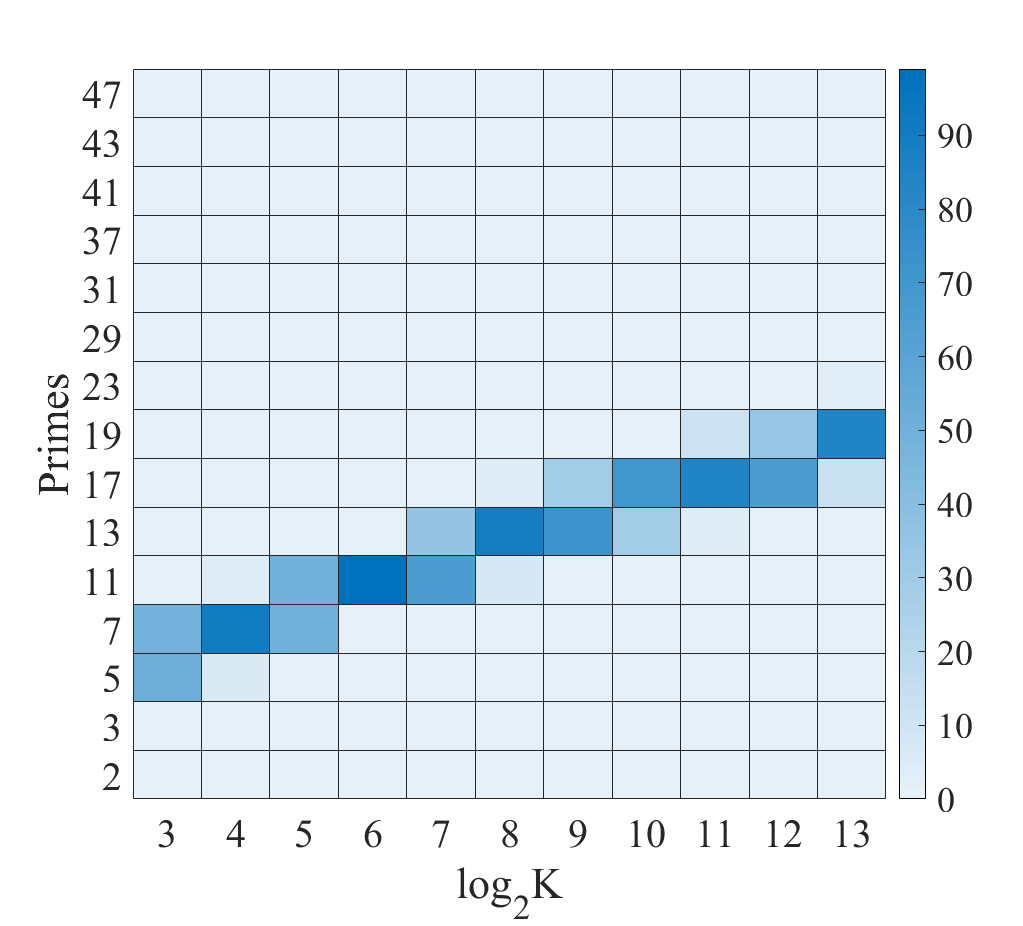}}                
\quad\quad\subfloat[]{\label{fig:l3_IC}\includegraphics[height=0.3\textwidth, width=.2\textwidth]{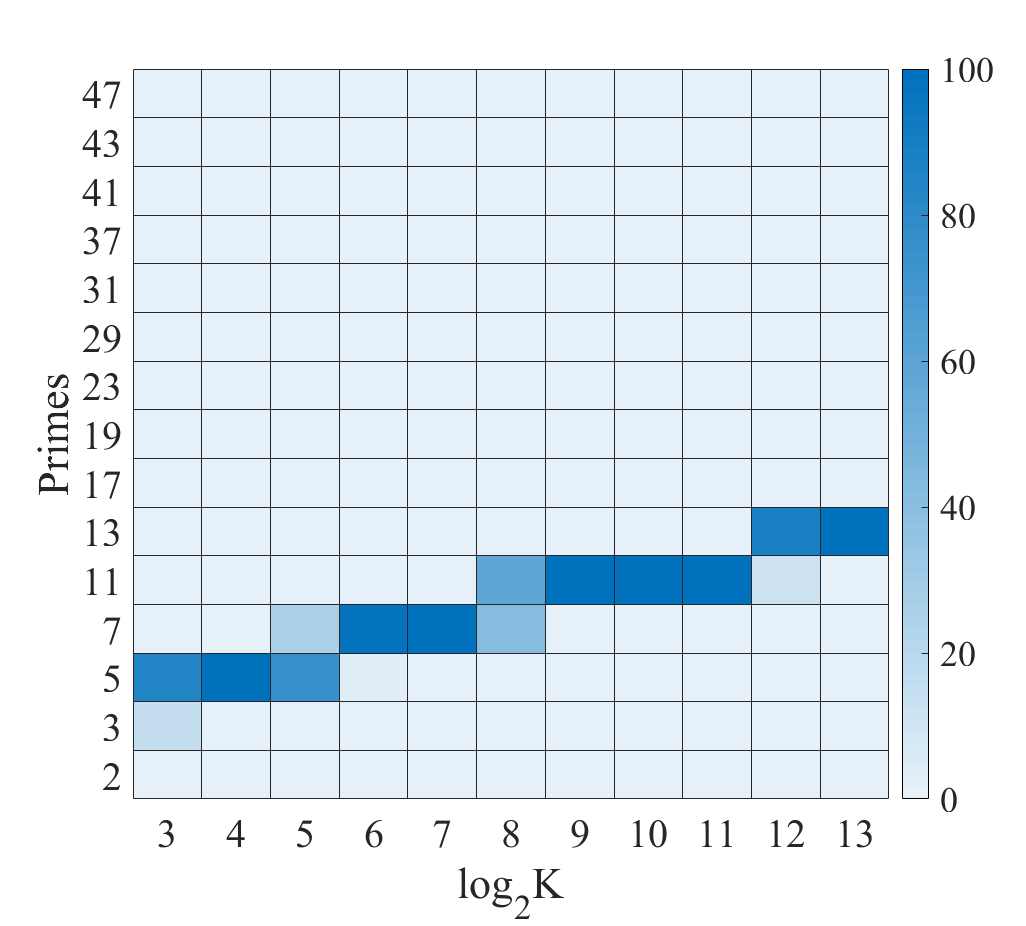}}
  \caption{L=3: Histogram for the primes per $\log_2{(K)}$ (a) without and (b) with interference cancellation.}
  \label{fig:l3}
\end{figure}

\begin{figure}[htb]
  \centering 
\subfloat[]{\label{}\includegraphics[height=0.3\textwidth, width=.2\textwidth]{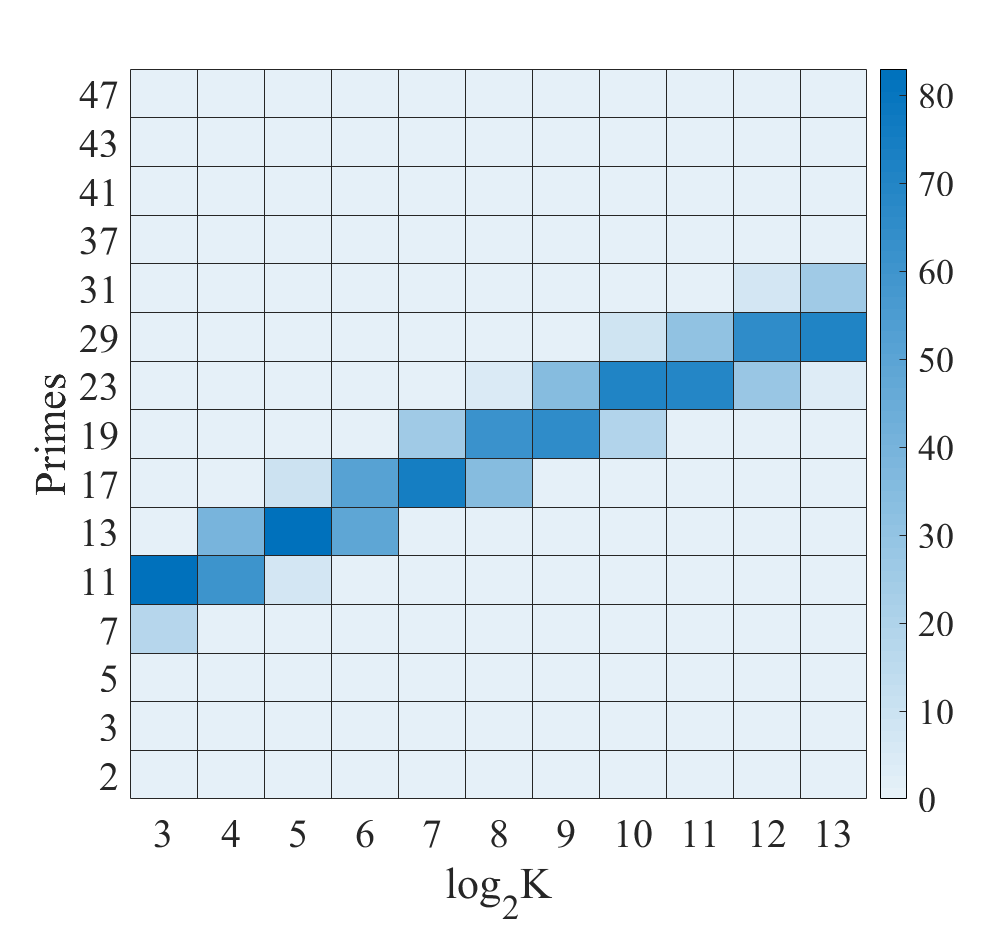}}                
\quad\quad\subfloat[]{\label{}\includegraphics[height=0.3\textwidth, width=.2\textwidth]{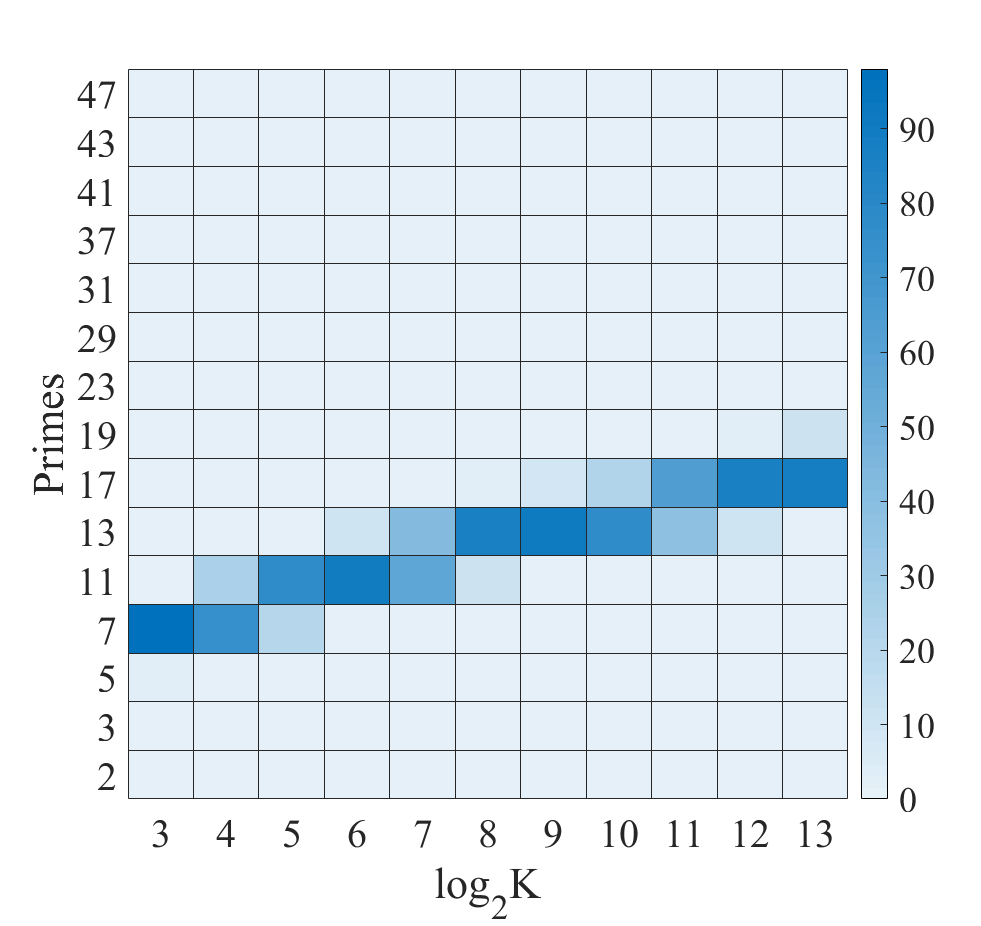}}
  \caption{L=5: Histogram for the primes per $\log_2{(K)}$ (a) without and (b) with interference cancellation.}
  \label{fig:l5}
\end{figure}

\begin{figure}[htb]
  \centering 
\subfloat[]{\label{}\includegraphics[height=0.3\textwidth, width=.2\textwidth]{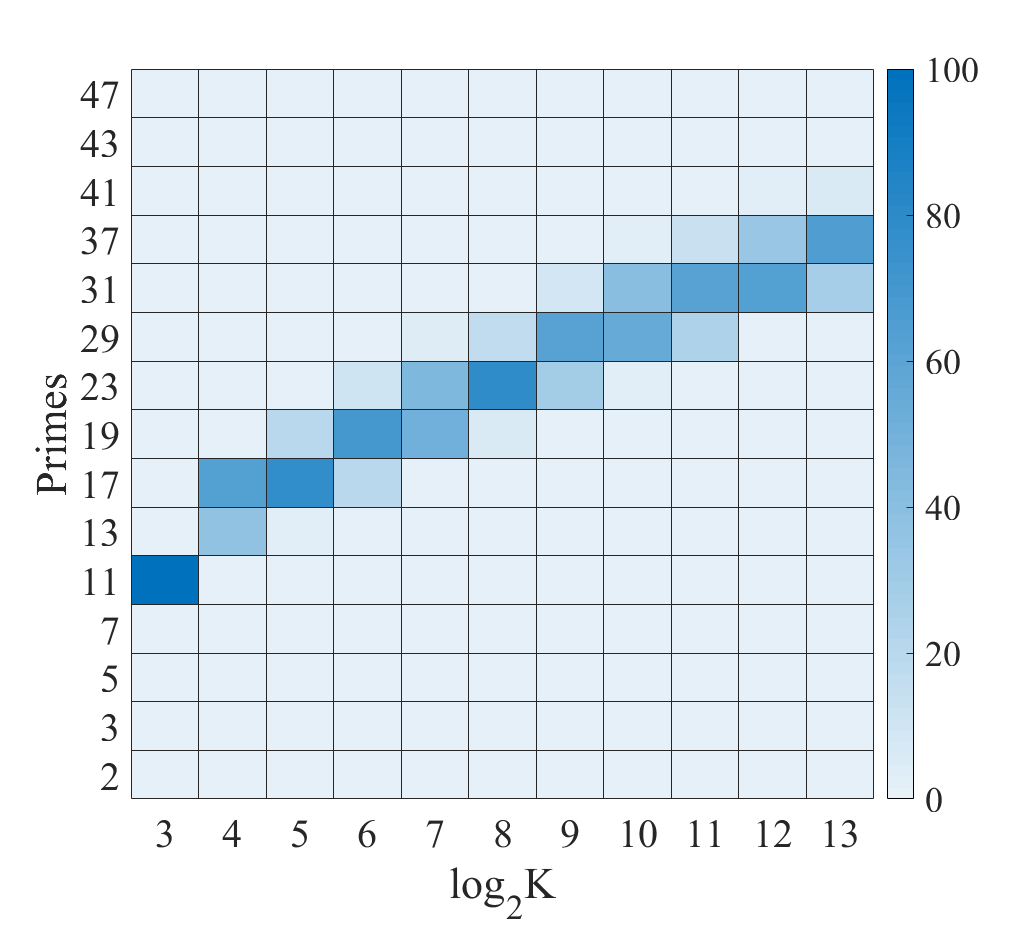}}                
\quad\quad\subfloat[]{\label{}\includegraphics[height=0.3\textwidth, width=.2\textwidth]{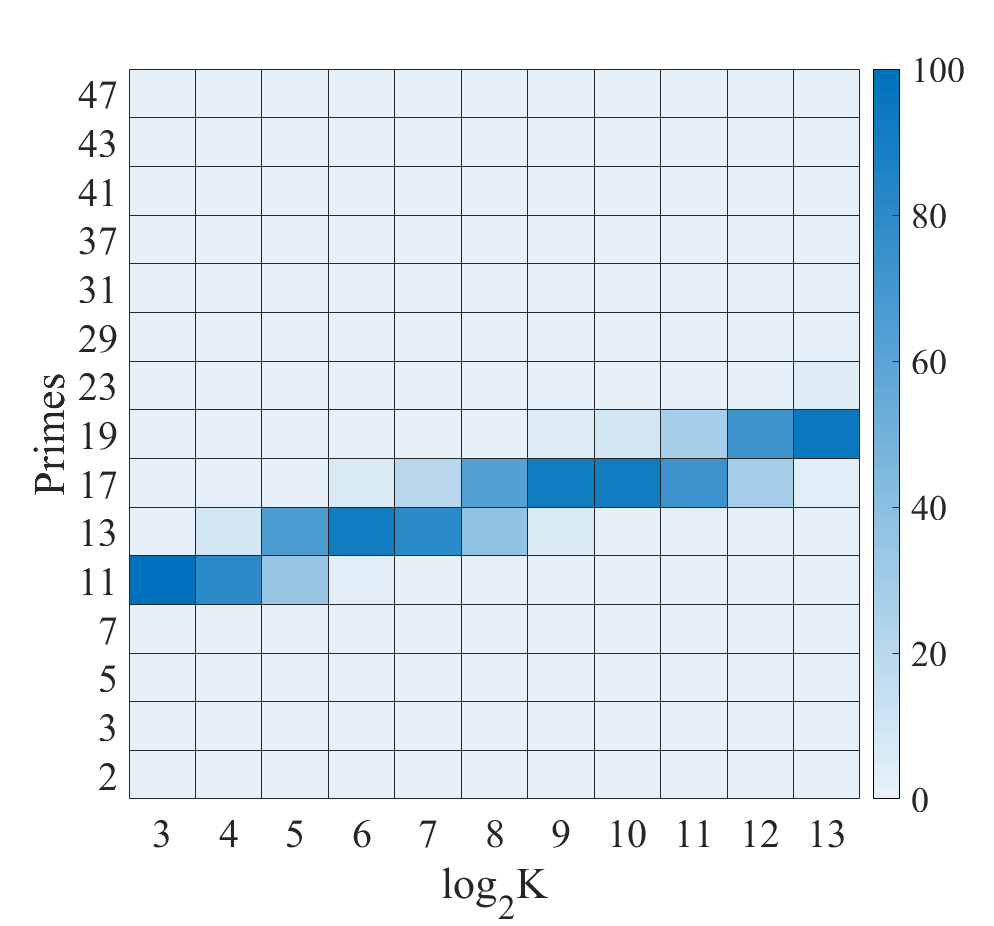}}
  \caption{L=7: Histogram for the primes per $\log_2{(K)}$ (a) without and (b) with interference cancellation.}
  \label{fig:l7}
\end{figure}

\section{Discussion and Concluding Remarks
}{
\label{sec:discussion}
In order to further investigate the discrepancy between the $\frac{\log^2(K)}{\log \log K}$ growth scale for the number of rounds suggested by the obtained theoretical upper bounds and the $\log(K)$ growth scale that is observed through simulations, we plot in Figures \ref{fig:l3}, \ref{fig:l5}, and \ref{fig:l7} the histograms of the prime numbers $p_i$ corresponding to the phase, during which, the algorithm terminates for different network sizes  with and without interference cancellation. We note how the prime number corresponding to the termination phase is typically significantly less than $L \log_2 K$, which we believe is the main reason for the above mentioned discrepancy and the looseness of the bound in Theorem \ref{thm1}. In future work, we plan to capitalize on this observation to derive a tighter upper bound and explain this phenomenon. 

We finally highlight the potential of the proposed algorithm by noting that in a locally connected network, where due to path loss constraints, each receiver $i$ can only be connected to transmitters with indices $j: |i-j|\leq r$ in a neighborhood of radius $r$, one can apply the proposed algorithm after replacing every transmitter index $j$ by $j \mod (2r+2)+1$, and then it is guaranteed that each receiver is connected to transmitters with distinct indices whose maximum value is at most $2r+1$, and our analysis and simulation results would hold with a replacement of the number of users $K$ by the diameter of the local connectivity neighborhood $2r+1$. This could lead to quite powerful results for very large networks with path loss constraints. In future work, we plan to demonstrate and investigate the practical effectiveness of the proposed algorithm via testing on dense a wireless sensor network testbed.


}
}

\bibliographystyle{IEEEtran}

\end{document}